\documentclass[10pt,a4paper,onecolumn]{article}
\usepackage{marginnote}
\usepackage{graphicx}
\usepackage{xcolor}
\usepackage{authblk,etoolbox}
\usepackage{titlesec}
\usepackage{calc}
\usepackage{tikz}
\usepackage{caption}
\usepackage{tcolorbox}
\usepackage{amssymb,amsmath}
\usepackage{ifxetex,ifluatex}
\usepackage{seqsplit}
\usepackage{xstring}
\usepackage[pdfa]{hyperref}
\usepackage{hyperxmp}
\hypersetup{unicode=true,
            pdfapart=3,
            pdfaconformance=B,
            pdftitle={POSEIDON: A Multidimensional Atmospheric Retrieval Code for Exoplanet Spectra},
            pdfauthor={Ryan J. MacDonald},
            pdfpublication={Journal of Open Source Software},
            pdfpublisher={Open Journals},
            pdfissn={2475-9066},
            pdfpubtype={journal},
            pdfvolumenum={8},
            pdfissuenum={81},
            pdfdoi={10.21105/joss.04873},
            pdfcopyright={Copyright (c) 2023, Ryan J. MacDonald},
            pdflicenseurl={http://creativecommons.org/licenses/by/4.0/},
            colorlinks=true,
            linkcolor=[rgb]{0.0, 0.5, 1.0},
            citecolor=Blue,
            urlcolor=[rgb]{0.0, 0.5, 1.0},
            breaklinks=true
}

\usepackage{float}
\let\origfigure\figure
\let\endorigfigure\endfigure
\renewenvironment{figure}[1][2] {
    \expandafter\origfigure\expandafter[H]
} {
    \endorigfigure
}

\usepackage{orcidlink}
\usepackage{fixltx2e} 
\usepackage[
  backend=biber,
]{biblatex}
\bibliography{POSEIDON.bib}{}

\let\textttOrig=\texttt
\def\texttt#1{\expandafter\textttOrig{\seqsplit{#1}}}
\renewcommand{\seqinsert}{\ifmmode
  \allowbreak
  \else\penalty6000\hspace{0pt plus 0.02em}\fi}

\makeatletter
\let\href@Orig=\href
\def\href@Urllike#1#2{\href@Orig{#1}{\begingroup
    \def\Url@String{#2}\Url@FormatString
    \endgroup}}
\def\href@Notdoi#1#2{\def\tempa{#1}\def\tempb{#2}%
  \ifx\tempa\tempb\relax\href@Urllike{#1}{#2}\else
  \href@Orig{#1}{#2}\fi}
\def\href#1#2{%
  \IfBeginWith{#1}{https://doi.org}%
  {\href@Urllike{#1}{#2}}{\href@Notdoi{#1}{#2}}}
\makeatother

\newlength{\cslhangindent}
\newlength{\csllabelwidth}
\newenvironment{CSLReferences}[3] 
 {
  \setlength{\parindent}{0pt}
  \ifodd #1 \everypar{\setlength{\hangindent}{\cslhangindent}}\ignorespaces\fi
  \ifnum #2 > 0
  \setlength{\parskip}{#2\baselineskip}
  \fi
 }%
 {}
\usepackage{calc}

\usepackage[top=3.5cm, bottom=3cm, right=1.5cm, left=1.0cm,
            headheight=2.2cm, reversemp, includemp, marginparwidth=4.5cm]{geometry}

\titleformat{\section}
  {\normalfont\sffamily\Large\bfseries}
  {}{0pt}{}
\titleformat{\subsection}
  {\normalfont\sffamily\large\bfseries}
  {}{0pt}{}
\titleformat{\subsubsection}
  {\normalfont\sffamily\bfseries}
  {}{0pt}{}
\titleformat*{\paragraph}
  {\sffamily\normalsize}

\usepackage{fancyhdr}
\makeatletter
\let\ps@plain\ps@fancy
\fancyheadoffset[L]{4.5cm}
\fancyfootoffset[L]{4.5cm}

\definecolor{linky}{rgb}{0.0, 0.5, 1.0}

\newtcolorbox{repobox}
   {colback=red, colframe=red!75!black,
     boxrule=0.5pt, arc=2pt, left=6pt, right=6pt, top=3pt, bottom=3pt}

\newcommand{\ExternalLink}{%
   \tikz[x=1.2ex, y=1.2ex, baseline=-0.05ex]{%
       \begin{scope}[x=1ex, y=1ex]
           \clip (-0.1,-0.1)
               --++ (-0, 1.2)
               --++ (0.6, 0)
               --++ (0, -0.6)
               --++ (0.6, 0)
               --++ (0, -1);
           \path[draw,
               line width = 0.5,
               rounded corners=0.5]
               (0,0) rectangle (1,1);
       \end{scope}
       \path[draw, line width = 0.5] (0.5, 0.5)
           -- (1, 1);
       \path[draw, line width = 0.5] (0.6, 1)
           -- (1, 1) -- (1, 0.6);
       }
   }

\patchcmd{\@maketitle}{center}{flushleft}{}{}
\patchcmd{\@maketitle}{center}{flushleft}{}{}
\patchcmd{\@maketitle}{\LARGE}{\LARGE\sffamily}{}{}
\def\maketitle{{%
  
  \AB@maketitle}}
\makeatletter
\renewcommand\AB@affilsepx{ \protect\Affilfont}
\renewcommand\AB@affilnote[1]{{\bfseries #1}\hspace{3pt}}
\renewcommand{\affil}[2][]%
   {\newaffiltrue\let\AB@blk@and\AB@pand
      \if\relax#1\relax\def\AB@note{\AB@thenote}\else\def\AB@note{#1}%
        \setcounter{Maxaffil}{0}\fi
        \begingroup
        \let\href=\href@Orig
        \let\texttt=\textttOrig
        \let\protect\@unexpandable@protect
        \def\thanks{\protect\thanks}\def\footnote{\protect\footnote}%
        \@temptokena=\expandafter{\AB@authors}%
        {\def\\{\protect\\\protect\Affilfont}\xdef\AB@temp{#2}}%
         \xdef\AB@authors{\the\@temptokena\AB@las\AB@au@str
         \protect\\[\affilsep]\protect\Affilfont\AB@temp}%
         \gdef\AB@las{}\gdef\AB@au@str{}%
        {\def\\{, \ignorespaces}\xdef\AB@temp{#2}}%
        \@temptokena=\expandafter{\AB@affillist}%
        \xdef\AB@affillist{\the\@temptokena \AB@affilsep
          \AB@affilnote{\AB@note}\protect\Affilfont\AB@temp}%
      \endgroup
       \let\AB@affilsep\AB@affilsepx
}
\makeatother

\renewcommand\Affilfont{\sffamily\small\mdseries}
\ifnum 0\ifxetex 1\fi\ifluatex 1\fi=0 
  \usepackage[T1]{fontenc}
  \usepackage[utf8]{inputenc}

\else 
  \ifxetex
    \usepackage{mathspec}
    \usepackage{fontspec}

  \else
    \usepackage{fontspec}
  \fi
  \defaultfontfeatures{Ligatures=TeX,Scale=MatchLowercase}

\fi
\IfFileExists{upquote.sty}{\usepackage{upquote}}{}
\IfFileExists{microtype.sty}{%
\usepackage{microtype}
\UseMicrotypeSet[protrusion]{basicmath} 
}{}

\let\addcontentslineOrig=\addcontentsline
\def\addcontentsline#1#2#3{\bgroup
  \let\texttt=\textttOrig\addcontentslineOrig{#1}{#2}{#3}\egroup}
\let\markbothOrig\markboth
\def\markboth#1#2{\bgroup
  \let\texttt=\textttOrig\markbothOrig{#1}{#2}\egroup}
\let\markrightOrig\markright
\def\markright#1{\bgroup
  \let\texttt=\textttOrig\markrightOrig{#1}\egroup}

\usepackage{graphicx,grffile}
\makeatletter
\def\maxwidth{\ifdim\Gin@nat@width>\linewidth\linewidth\else\Gin@nat@width\fi}
\def\maxheight{\ifdim\Gin@nat@height>\textheight\textheight\else\Gin@nat@height\fi}
\makeatother
\setkeys{Gin}{width=\maxwidth,height=\maxheight,keepaspectratio}
\IfFileExists{parskip.sty}{%
\usepackage{parskip}
}{
\setlength{\parindent}{0pt}
\setlength{\parskip}{6pt plus 2pt minus 1pt}
}
\ifx\paragraph\undefined\else
\let\oldparagraph\paragraph
\renewcommand{\paragraph}[1]{\oldparagraph{#1}\mbox{}}
\fi
\ifx\subparagraph\undefined\else
\let\oldsubparagraph\subparagraph
\renewcommand{\subparagraph}[1]{\oldsubparagraph{#1}\mbox{}}
\fi

\title{\texttt{POSEIDON}: A Multidimensional Atmospheric Retrieval Code for Exoplanet Spectra}

        \author[1, 2, 3]{Ryan J. MacDonald\,\orcidlink{0000-0003-4816-3469}\,}
    
      \affil[1]{Department of Astronomy, University of Michigan, 1085 S.
University Ave., Ann Arbor, MI 48109, USA}
      \affil[2]{NHFP Sagan Fellow}
      \affil[3]{Department of Astronomy and Carl Sagan Institute,
Cornell University, 122 Sciences Drive, Ithaca, NY 14853, USA}
  \date{\vspace{-7ex}}

\begin{document}
\maketitle

\marginpar{

  \begin{flushleft}
  \sffamily\small

  {\bfseries DOI:} \href{https://doi.org/10.21105/joss.04873}{\color{linky}{10.21105/joss.04873}}

  \vspace{2mm}

  {\bfseries Software}
  \begin{itemize}
    \setlength\itemsep{0em}
    \item \href{https://github.com/openjournals/joss-reviews/issues/4873}{\color{linky}{Review}} \ExternalLink
    \item \href{https://github.com/MartianColonist/POSEIDON}{\color{linky}{Repository}} \ExternalLink
    \item \href{https://doi.org/10.5281/zenodo.7530543}{\color{linky}{Archive}} \ExternalLink
  \end{itemize}

  \vspace{2mm}

  \par\noindent\hrulefill\par

  \vspace{2mm}

  {\bfseries Editor:} \href{https://dfm.io/}{Dan Foreman-Mackey}\ExternalLink\orcidlink{0000-0002-9328-5652} \\
  \vspace{1mm}
    {\bfseries Reviewers:}
  \begin{itemize}
  \setlength\itemsep{0em}
    \item \href{https://github.com/ideasrule}{@ideasrule}
    \item \href{https://github.com/LorenzoMugnai}{@LorenzoMugnai}
    \end{itemize}
    \vspace{2mm}

  {\bfseries Submitted:} 24 August 2022\\
  {\bfseries Published:} 13 January 2023

  \vspace{2mm}
  {\bfseries License}\\
  Authors of papers retain copyright and release the work under a Creative Commons Attribution 4.0 International License (\href{http://creativecommons.org/licenses/by/4.0/}{\color{linky}{CC BY 4.0}}).

    \vspace{4mm}
  {\bfseries In partnership with}\\
  \vspace{2mm}
  \includegraphics[width=4cm]{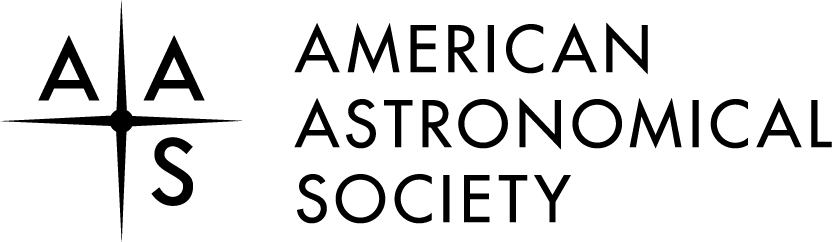}
  \vspace{2mm}
  \newline
  This article and software are linked with research article DOI \href{https://doi.org/10.3847/1538-4357/ac47fe}{\color{linky}{10.3847/1538-4357/ac47fe}}, published in the Astrophysical
Journal.

  \end{flushleft}
}

\vspace{0.5cm}

\hypertarget{summary}{%
\section{Summary}\label{summary}}

Exoplanet atmospheres are a dynamic and fast-changing field at the
frontier of modern astronomy. Telescope observations can reveal the
chemical composition, temperature, cloud properties, and (potentially)
the habitability of these remote worlds. Astronomers can measure these
atmospheric properties by observing how the fraction of starlight
blocked by a planet passing in front of its host star changes with
wavelength --- a technique called transmission spectroscopy. Since the
wavelengths where different atoms and molecules absorb are already known
(from laboratory measurements or quantum mechanics), astronomers can
compare models of exoplanet spectra to observations to infer the
chemical composition of exoplanets.

\texttt{POSEIDON} is a Python package for the modelling and analysis of
exoplanet spectra. \texttt{POSEIDON} has two main functions: (i)
computation of model spectra for 1D, 2D, or 3D exoplanet atmospheres;
and (ii) a Bayesian fitting routine (`atmospheric retrieval') that can
infer the range of atmospheric properties consistent with an observed
exoplanet spectrum.

\hypertarget{exoplanet-modelling-and-atmospheric-retrieval-with-poseidon}{%
\section{\texorpdfstring{Exoplanet Modelling and Atmospheric Retrieval with\\ 
\texttt{POSEIDON}}{Exoplanet Modelling and Atmospheric Retrieval with POSEIDON}}\label{exoplanet-modelling-and-atmospheric-retrieval-with-poseidon}}

\begin{figure}
\centering
\includegraphics[width=1\textwidth,height=\textheight]{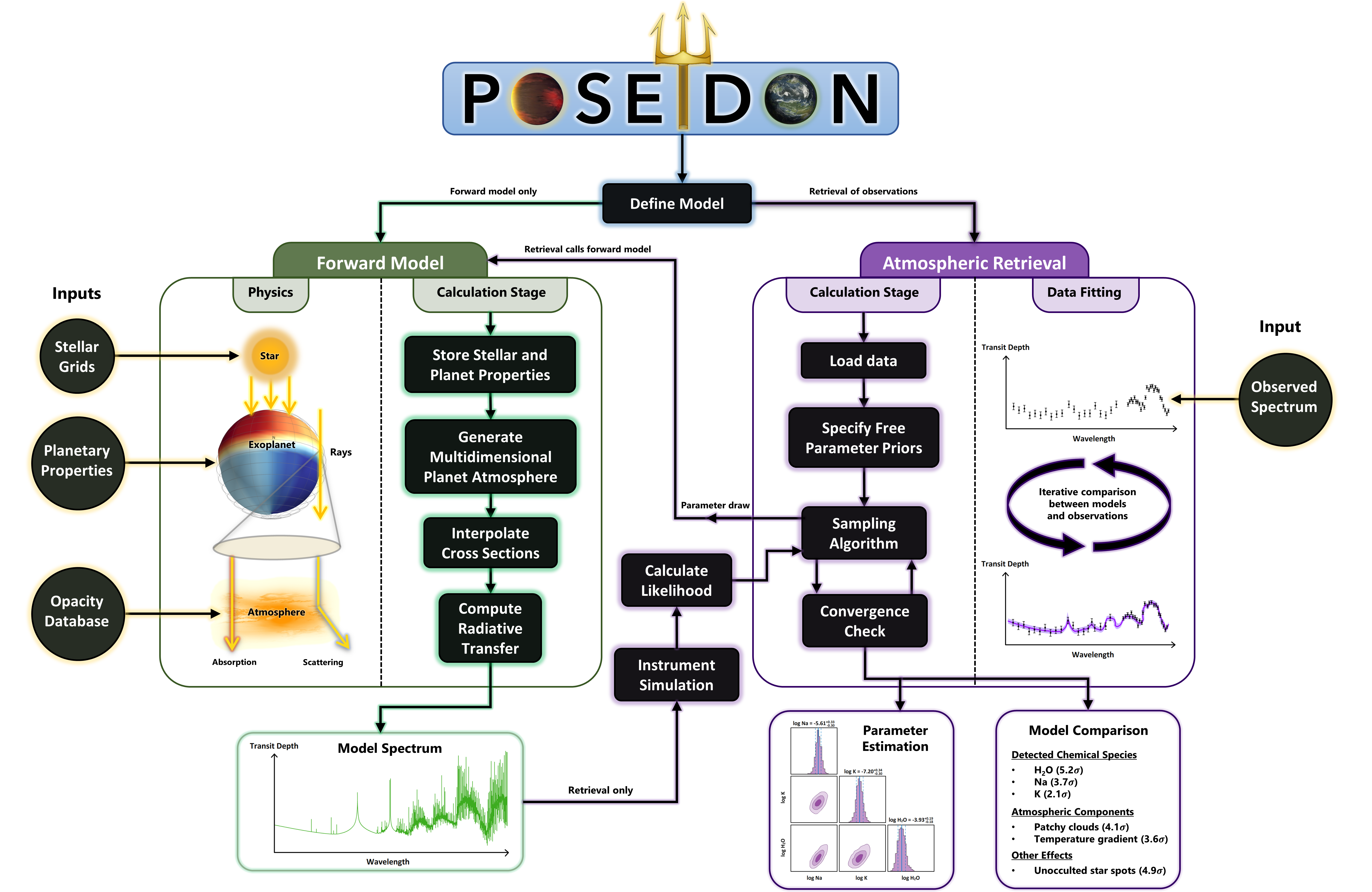}
\caption{Schematic architecture of the \texttt{POSEIDON} atmospheric
retrieval code. Users can call \texttt{POSEIDON} in two main ways: (i)
to generate a model exoplanet spectrum for a specified planet atmosphere
(green arrows); or (ii) to fit an observed exoplanet spectrum by
statistical sampling of a model's atmospheric properties (purple
arrows). The diagram highlights code inputs (circles), algorithm steps
(rectangles), and code outputs (bottom green or purple boxes).
\label{fig:POSEIDON_architecture}}
\end{figure}

The first major use case for \texttt{POSEIDON} is `forward modelling'
--- illustrated on the left of \autoref{fig:POSEIDON_architecture}. A
user can generate a model planet spectrum, for a given star-planet
system, by providing a specific set of atmospheric properties (e.g.~the
chemical composition and temperature). The forward model mode allows
users to explore how atmospheric properties alter an exoplanet spectrum
and to produce predicted model spectra for observing proposals. The
required input files (pre-computed stellar grids and an opacity
database) are available to download from an online repository (linked in
the documentation).

The second major use case for \texttt{POSEIDON} is atmospheric retrieval
--- illustrated on the right of \autoref{fig:POSEIDON_architecture}. To
initialise a retrieval, a user provides an observed exoplanet spectrum
and the range of atmospheric properties to be explored (i.e.~the prior
ranges for a set of free parameters defining a model). A Bayesian
statistical sampling algorithm --- nominally
\href{https://github.com/JohannesBuchner/PyMultiNest}{\texttt{PyMultiNest}}
(\protect\hyperlink{Buchner:2014}{Buchner et al., 2014}) --- then repeatedly calls the forward model,
comparing the generated spectrum to the observations, until the
parameter space is fully explored and a convergence criteria reached.
The main outputs of an atmospheric retrieval are the posterior
probability distributions of the model parameters and the model's
Bayesian evidence. The Bayesian evidences from multiple retrievals, in
turn, can be subsequently compared to compute a detection significance
for each model component (e.g.~the statistical confidence for a molecule
being present in the planetary atmosphere).

\texttt{POSEIDON} was first described in the exoplanet literature by
\protect\hyperlink{MacDonald:2017}{MacDonald \& Madhusudhan (2017)}. 
Since then, the code has been used in 17 peer-reviewed publications 
(e.g., \protect\hyperlink{Sedaghati:2017}{Sedaghati et al., 2017}; 
\protect\hyperlink{Kaltenegger:2020}{Kaltenegger \& MacDonald et al., 2020}; \protect\hyperlink{Alam:2021}{Alam et al., 2021}). 
Most recently, a detailed description of \texttt{POSEIDON}'s new multidimensional forward model,
\texttt{TRIDENT}, was provided by \protect\hyperlink{MacDonald:2022}{MacDonald \& Lewis (2022)}.

\hypertarget{statement-of-need}{%
\section{Statement of Need}\label{statement-of-need}}

Recent years have seen a substantial improvement in the number of
high-quality exoplanet spectra. In particular, the newly operational
JWST and a profusion of high-resolution ground-based spectrographs offer
an abundance of exoplanet data. The accurate interpretation of such data
requires a retrieval code that can rapidly explore complex parameter
spaces describing a rich variety of atmospheric phenomena.

\texttt{POSEIDON} provides the capability to model and retrieve
transmission spectra of planets with inhomogeneous temperatures,
compositions, and cloud properties (i.e.~2D or 3D models). Several
studies have highlighted that not including these multidimensional
effects can bias retrieval inferences (e.g., \protect\hyperlink{Line:2016}{Line \& Parmentier, 2016}; 
\protect\hyperlink{Caldas:2019}{Caldas et al., 2019}; \protect\hyperlink{MacDonald:2020}{MacDonald et al., 2020}; \protect\hyperlink{Pluriel:2022}{Pluriel et al., 2022}).
However, existing open-source exoplanet retrieval codes assume 1D
atmospheres for computational efficiency. \texttt{POSEIDON}, therefore,
offers an open-source implementation of state-of-the-art
multidimensional retrieval methods (see \protect\hyperlink{MacDonald:2022}{MacDonald \& Lewis, 2022}) to aid the interpretation of high-quality
exoplanet spectra.

In a 1D configuration, \texttt{POSEIDON} compares well with other
retrieval codes. When applied to Hubble Space Telescope observations,
\texttt{POSEIDON} produces consistent retrieval results with the ATMO
and NEMESIS retrieval codes (\protect\hyperlink{Lewis:2020}{Lewis et al., 2020}; 
\protect\hyperlink{Rathcke:2021}{Rathcke et al., 2021}).
Recently, \protect\hyperlink{Barstow:2022}{Barstow et al., (2022)} presented a comparison of five
exoplanet retrieval codes, including \texttt{POSEIDON}, which
demonstrated good agreement on simulated Ariel (\protect\hyperlink{Tinetti:2020}{Tinetti et al., 2020}) 
transmission spectra. \texttt{POSEIDON} also offers exceptional
computational performance: a single 1D forward model over a wavelength
range sufficient for JWST analyses takes 70 ms (see \protect\hyperlink{MacDonald:2022}{MacDonald \& Lewis, 2022}, Appendix D), while publication-quality 1D retrievals typically take an hour or less. \texttt{POSEIDON} also supports multi-core
retrievals via \texttt{PyMultiNest}'s MPI implementation, which achieves
a roughly linear speed-up in the number of cores. Therefore,
\texttt{POSEIDON} allows users to readily explore 1D retrievals on
personal laptops while scaling up to multidimensional retrievals on
modest clusters.

\hypertarget{future-developments}{%
\section{Future Developments}\label{future-developments}}

\texttt{POSEIDON} v1.0 officially supports the modelling and retrieval
of exoplanet transmission spectra in 1D, 2D, and 3D. The initial release
also includes a beta version of thermal emission spectra modelling and
retrieval (for cloud-free, 1D atmospheres, with no scattering), which
will be developed further in future releases. Suggestions for additional
features are more than welcome.

\hypertarget{documentation}{%
\section{Documentation}\label{documentation}}

Documentation for \texttt{POSEIDON}, with step-by-step tutorials
illustrating research applications, is available at
\url{https://poseidon-retrievals.readthedocs.io/en/latest/}.

\hypertarget{similar-tools}{%
\section{Similar Tools}\label{similar-tools}}

The following exoplanet retrieval codes are open source:
\href{https://github.com/ideasrule/platon}{\texttt{PLATON}} (\protect\hyperlink{Zhang:2019}{Zhang et al., 2019}; \protect\hyperlink{Zhang:2020}{Zhang et al., 2020}),
\href{https://gitlab.com/mauricemolli/petitRADTRANS}{\texttt{petitRADTRANS}}
(\protect\hyperlink{Molliere:2019}{Mollière et al., 2019}),
\href{https://github.com/mrline/CHIMERA}{\texttt{CHIMERA}} (\protect\hyperlink{Line:2013}{Line et al., 2013}),
\href{https://github.com/ucl-exoplanets/TauREx3_public}{\texttt{TauRex}}
(\protect\hyperlink{Waldmann:2015}{Waldmann et al., 2015}; \protect\hyperlink{Al-Refaie:2021}{Al-Refaie et al., 2021}),
\href{https://github.com/nemesiscode/radtrancode}{\texttt{NEMESIS}}
(\protect\hyperlink{Irwin:2008}{Irwin et al., 2008})
\href{https://github.com/pcubillos/pyratbay}{\texttt{Pyrat\ Bay}}
(\protect\hyperlink{Cubillos:2021}{Cubillos \& Blecic, 2021}), and
\href{https://github.com/exosports/BART}{\texttt{BART}} (\protect\hyperlink{Harrington:2022}{Harrington et al., 2022})

\hypertarget{acknowledgements}{%
\section{Acknowledgements}\label{acknowledgements}}

RJM expresses gratitude to the developers of many open source Python
packages used by \texttt{POSEIDON}, in particular \texttt{Numba} (\protect\hyperlink{Lam:2015}{Lam et al., 2015}), \texttt{numpy} (\protect\hyperlink{Harris:2020}{Harris et al., 2020}), \texttt{Matplotlib}
(\protect\hyperlink{Hunter:2007}{Hunter, 2007}), \texttt{SciPy} (\protect\hyperlink{Virtanen:2020}{Virtanen et al., 2020}), and \texttt{Spectres} (\protect\hyperlink{Carnall:2017}{Carnall, 2017}).

RJM acknowledges financial support from the UK's Science and Technology
Facilities Council (STFC) during the early development of
\texttt{POSEIDON} and support from NASA Grant 80NSSC20K0586 issued
through the James Webb Space Telescope Guaranteed Time Observer Program.
Most recently, RJM acknowledges support from NASA through the NASA
Hubble Fellowship grant HST-HF2-51513.001 awarded by the Space Telescope
Science Institute, which is operated by the Association of Universities
for Research in Astronomy, Inc., for NASA, under contract NAS5-26555.
RJM is especially grateful to Lorenzo Mugnai and Michael Zhang for
excellent and helpful referee reports, and to the editor, Dan
Foreman-Mackey, for his tireless efforts to encourage new people to join
the open source community in astronomy. RJM thanks Nikole Lewis, Ishan
Mishra, Jonathan Gomez Barrientos, John Kappelmeier, Antonia Peters,
Kath Landgren, and Ruizhe Wang for helpful discussions.

\hypertarget{references}{%
\section*{References}\label{references}}
\addcontentsline{toc}{section}{References}

\hypertarget{refs}{}
\begin{CSLReferences}{1}{0}
\leavevmode\hypertarget{Alam:2021}{}%
Alam, M. K., López-Morales, M., MacDonald, R. J., Nikolov, N., Kirk, J.,
Goyal, J. M., Sing, D. K., Wakeford, H. R., Rathcke, A. D., Deming, D.
L., Sanz-Forcada, J., Lewis, N. K., Barstow, J. K., Mikal-Evans, T., \&
Buchhave, L. A. (2021). {Evidence of a Clear Atmosphere for WASP-62b:
The Only Known Transiting Gas Giant in the JWST Continuous Viewing
Zone}. \emph{Astrophysical Journal Letters}, \emph{906}(2), L10.
\url{https://doi.org/10.3847/2041-8213/abd18e}

\leavevmode\hypertarget{Al-Refaie:2021}{}%
Al-Refaie, A. F., Changeat, Q., Waldmann, I. P., \& Tinetti, G. (2021).
{TauREx 3: A Fast, Dynamic, and Extendable Framework for Retrievals}.
\emph{Astrophysical Journal}, \emph{917}(1), 37.
\url{https://doi.org/10.3847/1538-4357/ac0252}

\leavevmode\hypertarget{Barstow:2022}{}%
Barstow, J. K., Changeat, Q., Chubb, K. L., Cubillos, P. E., Edwards,
B., MacDonald, R. J., Min, M., \& Waldmann, I. P. (2022). {A retrieval
challenge exercise for the Ariel mission}. \emph{Experimental
Astronomy}, \emph{53}(2), 447--471.
\url{https://doi.org/10.1007/s10686-021-09821-w}

\leavevmode\hypertarget{Buchner:2014}{}%
Buchner, J., Georgakakis, A., Nandra, K., Hsu, L., Rangel, C.,
Brightman, M., Merloni, A., Salvato, M., Donley, J., \& Kocevski, D.
(2014). {X-ray spectral modelling of the AGN obscuring region in the
CDFS: Bayesian model selection and catalogue}. \emph{Astronomy \&
Astrophysics}, \emph{564}, A125.
\url{https://doi.org/10.1051/0004-6361/201322971}

\leavevmode\hypertarget{Caldas:2019}{}%
Caldas, A., Leconte, J., Selsis, F., Waldmann, I. P., Bordé, P.,
Rocchetto, M., \& Charnay, B. (2019). {Effects of a fully 3D atmospheric
structure on exoplanet transmission spectra: retrieval biases due to
day-night temperature gradients}. \emph{Astronomy \& Astrophysics},
\emph{623}, A161. \url{https://doi.org/10.1051/0004-6361/201834384}

\leavevmode\hypertarget{Carnall:2017}{}%
Carnall, A. C. (2017). {SpectRes: A Fast Spectral Resampling Tool in
Python}. \emph{arXiv e-Prints}, arXiv:1705.05165.
\url{http://arxiv.org/abs/1705.05165}

\leavevmode\hypertarget{Cubillos:2021}{}%
Cubillos, P. E., \& Blecic, J. (2021). {The PYRAT BAY framework for
exoplanet atmospheric modelling: a population study of Hubble/WFC3
transmission spectra}. \emph{Monthly Notices of the Royal Astronomical
Society}, \emph{505}(2), 2675--2702.
\url{https://doi.org/10.1093/mnras/stab1405}

\leavevmode\hypertarget{Harrington:2022}{}%
Harrington, J., Himes, M. D., Cubillos, P. E., Blecic, J., Rojo, P. M.,
Challener, R. C., Lust, N. B., Bowman, M. O., Blumenthal, S. D.,
Dobbs-Dixon, I., Foster, A. S. D., Foster, A. J., Green, M. R., Loredo,
T. J., McIntyre, K. J., Stemm, M. M., \& Wright, D. C. (2022). {An
Open-source Bayesian Atmospheric Radiative Transfer (BART) Code. I.
Design, Tests, and Application to Exoplanet HD 189733b}. \emph{Planetary
Science Journal}, \emph{3}(4), 80.
\url{https://doi.org/10.3847/PSJ/ac3513}

\leavevmode\hypertarget{Harris:2020}{}%
Harris, C. R., Millman, K. J., van der Walt, S. J., Gommers, R.,
Virtanen, P., Cournapeau, D., Wieser, E., Taylor, J., Berg, S., Smith,
N. J., Kern, R., Picus, M., Hoyer, S., van Kerkwijk, M. H., Brett, M.,
Haldane, A., del Rio, J. F., Wiebe, M., Peterson, P., \ldots{} Oliphant,
T. E. (2020). {Array programming with NumPy}. \emph{Nature},
\emph{585}(7825), 357--362.
\url{https://doi.org/10.1038/s41586-020-2649-2}

\leavevmode\hypertarget{Hunter:2007}{}%
Hunter, J. D. (2007). {Matplotlib: A 2D Graphics Environment}.
\emph{Computing in Science and Engineering}, \emph{9}(3), 90--95.
\url{https://doi.org/10.1109/MCSE.2007.55}

\leavevmode\hypertarget{Irwin:2008}{}%
Irwin, P. G. J., Teanby, N. A., de Kok, R., Fletcher, L. N., Howett, C.
J. A., Tsang, C. C. C., Wilson, C. F., Calcutt, S. B., Nixon, C. A., \&
Parrish, P. D. (2008). {The NEMESIS planetary atmosphere radiative
transfer and retrieval tool}. \emph{Journal of Quantitative Spectroscopy
and Radiative Transfer}, \emph{109}, 1136--1150.
\url{https://doi.org/10.1016/j.jqsrt.2007.11.006}

\leavevmode\hypertarget{Kaltenegger:2020}{}%
Kaltenegger, L., MacDonald, R. J., Kozakis, T., Lewis, N. K., Mamajek,
E. E., McDowell, J. C., \& Vanderburg, A. (2020). {The White Dwarf
Opportunity: Robust Detections of Molecules in Earth-like Exoplanet
Atmospheres with the James Webb Space Telescope}. \emph{Astrophysical
Journal Letters}, \emph{901}(1), L1.
\url{https://doi.org/10.3847/2041-8213/aba9d3}

\leavevmode\hypertarget{Lam:2015}{}%
Lam, S. K., Pitrou, A., \& Seibert, S. (2015). Numba: A llvm-based
python jit compiler. \emph{Proceedings of the Second Workshop on the
LLVM Compiler Infrastructure in HPC}, 1--6.

\leavevmode\hypertarget{Lewis:2020}{}%
Lewis, N. K., Wakeford, H. R., MacDonald, R. J., Goyal, J. M., Sing, D.
K., Barstow, J., Powell, D., Kataria, T., Mishra, I., Marley, M. S.,
Batalha, N. E., Moses, J. I., Gao, P., Wilson, T. J., Chubb, K. L.,
Mikal-Evans, T., Nikolov, N., Pirzkal, N., Spake, J. J., \ldots{} Zhang,
X. (2020). {Into the UV: The Atmosphere of the Hot Jupiter HAT-P-41b
Revealed}. \emph{Astrophysical Journal Letters}, \emph{902}(1), L19.
\url{https://doi.org/10.3847/2041-8213/abb77f}

\leavevmode\hypertarget{Line:2016}{}%
Line, M. R., \& Parmentier, V. (2016). {The Influence of Nonuniform
Cloud Cover on Transit Transmission Spectra}. \emph{Astrophysical
Journal}, \emph{820}(1), 78.
\url{https://doi.org/10.3847/0004-637X/820/1/78}

\leavevmode\hypertarget{Line:2013}{}%
Line, M. R., Wolf, A. S., Zhang, X., Knutson, H., Kammer, J. A.,
Ellison, E., Deroo, P., Crisp, D., \& Yung, Y. L. (2013). {A Systematic
Retrieval Analysis of Secondary Eclipse Spectra. I. A Comparison of
Atmospheric Retrieval Techniques}. \emph{Astrophysical Journal},
\emph{775}(2), 137. \url{https://doi.org/10.1088/0004-637X/775/2/137}

\leavevmode\hypertarget{MacDonald:2020}{}%
MacDonald, R. J., Goyal, J. M., \& Lewis, N. K. (2020). {Why Is it So
Cold in Here? Explaining the Cold Temperatures Retrieved from
Transmission Spectra of Exoplanet Atmospheres}. \emph{Astrophysical
Journal Letters}, \emph{893}(2), L43.
\url{https://doi.org/10.3847/2041-8213/ab8238}

\leavevmode\hypertarget{MacDonald:2022}{}%
MacDonald, R. J., \& Lewis, N. K. (2022). {TRIDENT: A Rapid 3D
Radiative-transfer Model for Exoplanet Transmission Spectra}.
\emph{Astrophysical Journal}, \emph{929}(1), 20.
\url{https://doi.org/10.3847/1538-4357/ac47fe}

\leavevmode\hypertarget{MacDonald:2017}{}%
MacDonald, R. J., \& Madhusudhan, N. (2017). {HD 209458b in new light:
evidence of nitrogen chemistry, patchy clouds and sub-solar water}.
\emph{Monthly Notices of the Royal Astronomical Society}, \emph{469}(2),
1979--1996. \url{https://doi.org/10.1093/mnras/stx804}

\leavevmode\hypertarget{Molliere:2019}{}%
Mollière, P., Wardenier, J. P., van Boekel, R., Henning, Th.,
Molaverdikhani, K., \& Snellen, I. A. G. (2019). {petitRADTRANS. A
Python radiative transfer package for exoplanet characterization and
retrieval}. \emph{Astronomy \& Astrophysics}, \emph{627}, A67.
\url{https://doi.org/10.1051/0004-6361/201935470}

\leavevmode\hypertarget{Pluriel:2022}{}%
Pluriel, W., Leconte, J., Parmentier, V., Zingales, T., Falco, A.,
Selsis, F., \& Bordé, P. (2022). {Toward a multidimensional analysis of
transmission spectroscopy. II. Day-night-induced biases in retrievals
from hot to ultrahot Jupiters}. \emph{Astronomy \& Astrophysics},
\emph{658}, A42. \url{https://doi.org/10.1051/0004-6361/202141943}

\leavevmode\hypertarget{Rathcke:2021}{}%
Rathcke, A. D., MacDonald, R. J., Barstow, J. K., Goyal, J. M.,
Lopez-Morales, M., Mendonça, J. M., Sanz-Forcada, J., Henry, G. W.,
Sing, D. K., Alam, M. K., Lewis, N. K., Chubb, K. L., Taylor, J.,
Nikolov, N., \& Buchhave, L. A. (2021). {HST PanCET Program: A Complete
Near-UV to Infrared Transmission Spectrum for the Hot Jupiter WASP-79b}.
\emph{Astronomical Journal}, \emph{162}(4), 138.
\url{https://doi.org/10.3847/1538-3881/ac0e99}

\leavevmode\hypertarget{Sedaghati:2017}{}%
Sedaghati, E., Boffin, H. M. J., MacDonald, R. J., Gandhi, S.,
Madhusudhan, N., Gibson, N. P., Oshagh, M., Claret, A., \& Rauer, H.
(2017). {Detection of titanium oxide in the atmosphere of a hot
Jupiter}. \emph{Nature}, \emph{549}(7671), 238--241.
\url{https://doi.org/10.1038/nature23651}

\leavevmode\hypertarget{Tinetti:2020}{}%
Tinetti, G., Eccleston, P., Haswell, C., Lagage, P.-O., Leconte, J.,
Lüftinger, T., Micela, G., Min, M., Pilbratt, G., Puig, L., \& al., et.
(2020). {Ariel: Enabling planetary science across light-years}.
\emph{ESA Ariel Mission Definition Study Report}, arXiv:2104.04824.
\url{https://doi.org/10.48550/arXiv.2104.04824}

\leavevmode\hypertarget{Virtanen:2020}{}%
Virtanen, P., Gommers, R., Oliphant, T. E., Haberland, M., Reddy, T.,
Cournapeau, D., Burovski, E., Peterson, P., Weckesser, W., Bright, J.,
van der Walt, S. J., Brett, M., Wilson, J., Millman, K. J., Mayorov, N.,
Nelson, A. R. J., Jones, E., Kern, R., Larson, E., \ldots{} SciPy 1. 0
Contributors. (2020). {SciPy 1.0: fundamental algorithms for scientific
computing in Python}. \emph{Nature Methods}, \emph{17}, 261--272.
\url{https://doi.org/10.1038/s41592-019-0686-2}

\leavevmode\hypertarget{Waldmann:2015}{}%
Waldmann, I. P., Tinetti, G., Rocchetto, M., Barton, E. J., Yurchenko,
S. N., \& Tennyson, J. (2015). {Tau-REx I: A Next Generation Retrieval
Code for Exoplanetary Atmospheres}. \emph{Astrophysical Journal},
\emph{802}(2), 107. \url{https://doi.org/10.1088/0004-637X/802/2/107}

\leavevmode\hypertarget{Zhang:2019}{}%
Zhang, M., Chachan, Y., Kempton, E. M.-R., \& Knutson, H. A. (2019).
{Forward Modeling and Retrievals with PLATON, a Fast Open-source Tool}.
\emph{Publications of the Astronomical Society of the Pacific},
\emph{131}(997), 034501. \url{https://doi.org/10.1088/1538-3873/aaf5ad}

\leavevmode\hypertarget{Zhang:2020}{}%
Zhang, M., Chachan, Y., Kempton, E. M.-R., Knutson, H. A., \& Chang, W.
(Happy). (2020). {PLATON II: New Capabilities and a Comprehensive
Retrieval on HD 189733b Transit and Eclipse Data}. \emph{Astrophysical
Journal}, \emph{899}(1), 27.
\url{https://doi.org/10.3847/1538-4357/aba1e6}

\end{CSLReferences}


@ARTICLE{MacDonald:2017,
       author = {{MacDonald}, Ryan J. and {Madhusudhan}, Nikku},
        title = "{HD 209458b in new light: evidence of nitrogen chemistry, patchy clouds and sub-solar water}",
      journal = {Monthly Notices of the Royal Astronomical Society},
         year = "2017",
        month = "Aug",
       volume = {469},
       number = {2},
        pages = {1979-1996},
          doi = {10.1093/mnras/stx804},
archivePrefix = {arXiv},
       eprint = {1701.01113},
 primaryClass = {astro-ph.EP},
       adsurl = {https://ui.adsabs.harvard.edu/abs/2017MNRAS.469.1979M}
}

@ARTICLE{MacDonald:2022,
       author = {{MacDonald}, Ryan J. and {Lewis}, Nikole K.},
        title = "{TRIDENT: A Rapid 3D Radiative-transfer Model for Exoplanet Transmission Spectra}",
      journal = {Astrophysical Journal},
         year = 2022,
        month = apr,
       volume = {929},
       number = {1},
          eid = {20},
        pages = {20},
          doi = {10.3847/1538-4357/ac47fe},
archivePrefix = {arXiv},
       eprint = {2111.05862},
 primaryClass = {astro-ph.EP},
       adsurl = {https://ui.adsabs.harvard.edu/abs/2022ApJ...929...20M}
}

@ARTICLE{Zhang:2019,
       author = {{Zhang}, Michael and {Chachan}, Yayaati and {Kempton}, Eliza M. -R. and {Knutson}, Heather A.},
        title = "{Forward Modeling and Retrievals with PLATON, a Fast Open-source Tool}",
      journal = {Publications of the Astronomical Society of the Pacific},
         year = 2019,
        month = mar,
       volume = {131},
       number = {997},
        pages = {034501},
          doi = {10.1088/1538-3873/aaf5ad},
archivePrefix = {arXiv},
       eprint = {1811.11761},
 primaryClass = {astro-ph.EP},
       adsurl = {https://ui.adsabs.harvard.edu/abs/2019PASP..131c4501Z}
}

@ARTICLE{Zhang:2020,
       author = {{Zhang}, Michael and {Chachan}, Yayaati and {Kempton}, Eliza M. -R. and {Knutson}, Heather A. and {Chang}, Wenjun (Happy)},
        title = "{PLATON II: New Capabilities and a Comprehensive Retrieval on HD 189733b Transit and Eclipse Data}",
      journal = {Astrophysical Journal},
         year = 2020,
        month = aug,
       volume = {899},
       number = {1},
          eid = {27},
        pages = {27},
          doi = {10.3847/1538-4357/aba1e6},
archivePrefix = {arXiv},
       eprint = {2004.09513},
 primaryClass = {astro-ph.EP},
       adsurl = {https://ui.adsabs.harvard.edu/abs/2020ApJ...899...27Z}
}

@ARTICLE{Molliere:2019,
       author = {{Molli{\`e}re}, P. and {Wardenier}, J.~P. and {van Boekel}, R. and {Henning}, Th. and {Molaverdikhani}, K. and {Snellen}, I.~A.~G.},
        title = "{petitRADTRANS. A Python radiative transfer package for exoplanet characterization and retrieval}",
      journal = {Astronomy \& Astrophysics},
         year = 2019,
        month = jul,
       volume = {627},
          eid = {A67},
        pages = {A67},
          doi = {10.1051/0004-6361/201935470},
archivePrefix = {arXiv},
       eprint = {1904.11504},
 primaryClass = {astro-ph.EP},
       adsurl = {https://ui.adsabs.harvard.edu/abs/2019A&A...627A..67M}
}

@ARTICLE{Line:2013,
       author = {{Line}, Michael R. and {Wolf}, Aaron S. and {Zhang}, Xi and {Knutson}, Heather and {Kammer}, Joshua A. and {Ellison}, Elias and {Deroo}, Pieter and {Crisp}, Dave and {Yung}, Yuk L.},
        title = "{A Systematic Retrieval Analysis of Secondary Eclipse Spectra. I. A Comparison of Atmospheric Retrieval Techniques}",
      journal = {Astrophysical Journal},
         year = 2013,
        month = oct,
       volume = {775},
       number = {2},
          eid = {137},
        pages = {137},
          doi = {10.1088/0004-637X/775/2/137},
archivePrefix = {arXiv},
       eprint = {1304.5561},
 primaryClass = {astro-ph.EP},
       adsurl = {https://ui.adsabs.harvard.edu/abs/2013ApJ...775..137L}
}

@ARTICLE{Cubillos:2021,
       author = {{Cubillos}, Patricio E. and {Blecic}, Jasmina},
        title = "{The PYRAT BAY framework for exoplanet atmospheric modelling: a population study of Hubble/WFC3 transmission spectra}",
      journal = {Monthly Notices of the Royal Astronomical Society},
         year = 2021,
        month = aug,
       volume = {505},
       number = {2},
        pages = {2675-2702},
          doi = {10.1093/mnras/stab1405},
archivePrefix = {arXiv},
       eprint = {2105.05598},
 primaryClass = {astro-ph.EP},
       adsurl = {https://ui.adsabs.harvard.edu/abs/2021MNRAS.505.2675C}
}

@ARTICLE{Waldmann:2015,
       author = {{Waldmann}, I.~P. and {Tinetti}, G. and {Rocchetto}, M. and {Barton}, E.~J. and {Yurchenko}, S.~N. and {Tennyson}, J.},
        title = "{Tau-REx I: A Next Generation Retrieval Code for Exoplanetary Atmospheres}",
      journal = {Astrophysical Journal},
         year = 2015,
        month = apr,
       volume = {802},
       number = {2},
          eid = {107},
        pages = {107},
          doi = {10.1088/0004-637X/802/2/107},
archivePrefix = {arXiv},
       eprint = {1409.2312},
 primaryClass = {astro-ph.EP},
       adsurl = {https://ui.adsabs.harvard.edu/abs/2015ApJ...802..107W}
}

@ARTICLE{Al-Refaie:2021,
       author = {{Al-Refaie}, A.~F. and {Changeat}, Q. and {Waldmann}, I.~P. and {Tinetti}, G.},
        title = "{TauREx 3: A Fast, Dynamic, and Extendable Framework for Retrievals}",
      journal = {Astrophysical Journal},
         year = 2021,
        month = aug,
       volume = {917},
       number = {1},
          eid = {37},
        pages = {37},
          doi = {10.3847/1538-4357/ac0252},
archivePrefix = {arXiv},
       eprint = {1912.07759},
 primaryClass = {astro-ph.IM},
       adsurl = {https://ui.adsabs.harvard.edu/abs/2021ApJ...917...37A}
}

@ARTICLE{Harrington:2022,
       author = {{Harrington}, Joseph and {Himes}, Michael D. and {Cubillos}, Patricio E. and {Blecic}, Jasmina and {Rojo}, Patricio M. and {Challener}, Ryan C. and {Lust}, Nate B. and {Bowman}, M. Oliver and {Blumenthal}, Sarah D. and {Dobbs-Dixon}, Ian and {Foster}, Andrew S.~D. and {Foster}, Austin J. and {Green}, M.~R. and {Loredo}, Thomas J. and {McIntyre}, Kathleen J. and {Stemm}, Madison M. and {Wright}, David C.},
        title = "{An Open-source Bayesian Atmospheric Radiative Transfer (BART) Code. I. Design, Tests, and Application to Exoplanet HD 189733b}",
      journal = {Planetary science Journal},
         year = 2022,
        month = apr,
       volume = {3},
       number = {4},
          eid = {80},
        pages = {80},
          doi = {10.3847/PSJ/ac3513},
archivePrefix = {arXiv},
       eprint = {2104.12522},
 primaryClass = {astro-ph.EP},
       adsurl = {https://ui.adsabs.harvard.edu/abs/2022PSJ.....3...80H}
}

@inproceedings{Lam:2015,
  title={Numba: A llvm-based python jit compiler},
  author={Lam, Siu Kwan and Pitrou, Antoine and Seibert, Stanley},
  booktitle={Proceedings of the Second Workshop on the LLVM Compiler Infrastructure in HPC},
  pages={1--6},
  year={2015}
}

@ARTICLE{Carnall:2017,
       author = {{Carnall}, A.~C.},
        title = "{SpectRes: A Fast Spectral Resampling Tool in Python}",
      journal = {arXiv e-prints},
         year = 2017,
        month = may,
          eid = {arXiv:1705.05165},
        pages = {arXiv:1705.05165},
archivePrefix = {arXiv},
       eprint = {1705.05165},
 primaryClass = {astro-ph.IM},
       adsurl = {https://ui.adsabs.harvard.edu/abs/2017arXiv170505165C}
}

@ARTICLE{Hunter:2007,
       author = {{Hunter}, John D.},
        title = "{Matplotlib: A 2D Graphics Environment}",
      journal = {Computing in Science and Engineering},
         year = 2007,
        month = may,
       volume = {9},
       number = {3},
        pages = {90-95},
          doi = {10.1109/MCSE.2007.55},
       adsurl = {https://ui.adsabs.harvard.edu/abs/2007CSE.....9...90H}
}

@ARTICLE{Harris:2020,
       author = {{Harris}, Charles R. and {Millman}, K. Jarrod and {van der Walt}, St{\'e}fan J. and {Gommers}, Ralf and {Virtanen}, Pauli and {Cournapeau}, David and {Wieser}, Eric and {Taylor}, Julian and {Berg}, Sebastian and {Smith}, Nathaniel J. and {Kern}, Robert and {Picus}, Matti and {Hoyer}, Stephan and {van Kerkwijk}, Marten H. and {Brett}, Matthew and {Haldane}, Allan and {del R{\'\i}o}, Jaime Fern{\'a}ndez and {Wiebe}, Mark and {Peterson}, Pearu and {G{\'e}rard-Marchant}, Pierre and {Sheppard}, Kevin and {Reddy}, Tyler and {Weckesser}, Warren and {Abbasi}, Hameer and {Gohlke}, Christoph and {Oliphant}, Travis E.},
        title = "{Array programming with NumPy}",
      journal = {Nature},
         year = 2020,
        month = sep,
       volume = {585},
       number = {7825},
        pages = {357-362},
          doi = {10.1038/s41586-020-2649-2},
archivePrefix = {arXiv},
       eprint = {2006.10256},
 primaryClass = {cs.MS},
       adsurl = {https://ui.adsabs.harvard.edu/abs/2020Natur.585..357H}
}

@ARTICLE{Virtanen:2020,
       author = {{Virtanen}, Pauli and {Gommers}, Ralf and {Oliphant}, Travis E. and {Haberland}, Matt and {Reddy}, Tyler and {Cournapeau}, David and {Burovski}, Evgeni and {Peterson}, Pearu and {Weckesser}, Warren and {Bright}, Jonathan and {van der Walt}, St{\'e}fan J. and {Brett}, Matthew and {Wilson}, Joshua and {Millman}, K. Jarrod and {Mayorov}, Nikolay and {Nelson}, Andrew R.~J. and {Jones}, Eric and {Kern}, Robert and {Larson}, Eric and {Carey}, C.~J. and {Polat}, {\.I}lhan and {Feng}, Yu and {Moore}, Eric W. and {VanderPlas}, Jake and {Laxalde}, Denis and {Perktold}, Josef and {Cimrman}, Robert and {Henriksen}, Ian and {Quintero}, E.~A. and {Harris}, Charles R. and {Archibald}, Anne M. and {Ribeiro}, Ant{\^o}nio H. and {Pedregosa}, Fabian and {van Mulbregt}, Paul and {SciPy 1. 0 Contributors}},
        title = "{SciPy 1.0: fundamental algorithms for scientific computing in Python}",
      journal = {Nature Methods},
         year = 2020,
        month = feb,
       volume = {17},
        pages = {261-272},
          doi = {10.1038/s41592-019-0686-2},
archivePrefix = {arXiv},
       eprint = {1907.10121},
 primaryClass = {cs.MS},
       adsurl = {https://ui.adsabs.harvard.edu/abs/2020NatMe..17..261V}
}

@ARTICLE{Buchner:2014,
       author = {{Buchner}, J. and {Georgakakis}, A. and {Nandra}, K. and {Hsu}, L. and
         {Rangel}, C. and {Brightman}, M. and {Merloni}, A. and {Salvato}, M. and
         {Donley}, J. and {Kocevski}, D.},
        title = "{X-ray spectral modelling of the AGN obscuring region in the CDFS: Bayesian model selection and catalogue}",
      journal = {Astronomy \& Astrophysics},
         year = "2014",
        month = "Apr",
       volume = {564},
          eid = {A125},
        pages = {A125},
          doi = {10.1051/0004-6361/201322971},
archivePrefix = {arXiv},
       eprint = {1402.0004},
 primaryClass = {astro-ph.HE},
       adsurl = {https://ui.adsabs.harvard.edu/\#abs/2014Astronomy and Astrophysics...564A.125B}
}

@ARTICLE{Sedaghati:2017,
       author = {{Sedaghati}, Elyar and {Boffin}, Henri M.~J. and {MacDonald}, Ryan J. and {Gandhi}, Siddharth and {Madhusudhan}, Nikku and {Gibson}, Neale P. and {Oshagh}, Mahmoudreza and {Claret}, Antonio and {Rauer}, Heike},
        title = "{Detection of titanium oxide in the atmosphere of a hot Jupiter}",
      journal = {Nature},
         year = 2017,
        month = sep,
       volume = {549},
       number = {7671},
        pages = {238-241},
          doi = {10.1038/nature23651},
archivePrefix = {arXiv},
       eprint = {1709.04118},
 primaryClass = {astro-ph.EP},
       adsurl = {https://ui.adsabs.harvard.edu/abs/2017Natur.549..238S}
}

@ARTICLE{Kaltenegger:2020,
       author = {{Kaltenegger}, Lisa and {MacDonald}, Ryan J. and {Kozakis}, Thea and {Lewis}, Nikole K. and {Mamajek}, Eric E. and {McDowell}, Jonathan C. and {Vanderburg}, Andrew},
        title = "{The White Dwarf Opportunity: Robust Detections of Molecules in Earth-like Exoplanet Atmospheres with the James Webb Space Telescope}",
      journal = {Astrophysical Journal Letters},
         year = 2020,
        month = sep,
       volume = {901},
       number = {1},
          eid = {L1},
        pages = {L1},
          doi = {10.3847/2041-8213/aba9d3},
archivePrefix = {arXiv},
       eprint = {2009.07274},
 primaryClass = {astro-ph.EP},
       adsurl = {https://ui.adsabs.harvard.edu/abs/2020ApJ...901L...1K}
}

@ARTICLE{Alam:2021,
       author = {{Alam}, Munazza K. and {L{\'o}pez-Morales}, Mercedes and {MacDonald}, Ryan J. and {Nikolov}, Nikolay and {Kirk}, James and {Goyal}, Jayesh M. and {Sing}, David K. and {Wakeford}, Hannah R. and {Rathcke}, Alexander D. and {Deming}, Drake L. and {Sanz-Forcada}, Jorge and {Lewis}, Nikole K. and {Barstow}, Joanna K. and {Mikal-Evans}, Thomas and {Buchhave}, Lars A.},
        title = "{Evidence of a Clear Atmosphere for WASP-62b: The Only Known Transiting Gas Giant in the JWST Continuous Viewing Zone}",
      journal = {Astrophysical Journal Letters},
         year = 2021,
        month = jan,
       volume = {906},
       number = {2},
          eid = {L10},
        pages = {L10},
          doi = {10.3847/2041-8213/abd18e},
archivePrefix = {arXiv},
       eprint = {2011.06424},
 primaryClass = {astro-ph.EP},
       adsurl = {https://ui.adsabs.harvard.edu/abs/2021ApJ...906L..10A}
}

@ARTICLE{Line:2016,
       author = {{Line}, Michael R. and {Parmentier}, Vivien},
        title = "{The Influence of Nonuniform Cloud Cover on Transit Transmission Spectra}",
      journal = {Astrophysical Journal},
         year = "2016",
        month = "Mar",
       volume = {820},
       number = {1},
          eid = {78},
        pages = {78},
          doi = {10.3847/0004-637X/820/1/78},
archivePrefix = {arXiv},
       eprint = {1511.09443},
 primaryClass = {astro-ph.EP},
       adsurl = {https://ui.adsabs.harvard.edu/abs/2016ApJ...820...78L}
}

@ARTICLE{Caldas:2019,
       author = {{Caldas}, A. and {Leconte}, J. and {Selsis}, F. and {Waldmann}, I.~P. and
         {Bord{\'e}}, P. and {Rocchetto}, M. and {Charnay}, B.},
        title = "{Effects of a fully 3D atmospheric structure on exoplanet transmission spectra: retrieval biases due to day-night temperature gradients}",
      journal = {Astronomy \& Astrophysics},
         year = "2019",
        month = "Mar",
       volume = {623},
          eid = {A161},
        pages = {A161},
          doi = {10.1051/0004-6361/201834384},
archivePrefix = {arXiv},
       eprint = {1901.09932},
 primaryClass = {astro-ph.EP},
       adsurl = {https://ui.adsabs.harvard.edu/abs/2019A&A...623A.161C}
}

@ARTICLE{MacDonald:2020,
       author = {{MacDonald}, Ryan J. and {Goyal}, Jayesh M. and {Lewis}, Nikole K.},
        title = "{Why Is it So Cold in Here? Explaining the Cold Temperatures Retrieved from Transmission Spectra of Exoplanet Atmospheres}",
      journal = {Astrophysical Journal Letters},
         year = 2020,
        month = apr,
       volume = {893},
       number = {2},
          eid = {L43},
        pages = {L43},
          doi = {10.3847/2041-8213/ab8238},
archivePrefix = {arXiv},
       eprint = {2003.11548},
 primaryClass = {astro-ph.EP},
       adsurl = {https://ui.adsabs.harvard.edu/abs/2020ApJ...893L..43M}
}

@ARTICLE{Pluriel:2022,
       author = {{Pluriel}, William and {Leconte}, J{\'e}r{\'e}my and {Parmentier}, Vivien and {Zingales}, Tiziano and {Falco}, Aur{\'e}lien and {Selsis}, Franck and {Bord{\'e}}, Pascal},
        title = "{Toward a multidimensional analysis of transmission spectroscopy. II. Day-night-induced biases in retrievals from hot to ultrahot Jupiters}",
      journal = {Astronomy \& Astrophysics},
         year = 2022,
        month = feb,
       volume = {658},
          eid = {A42},
        pages = {A42},
          doi = {10.1051/0004-6361/202141943},
archivePrefix = {arXiv},
       eprint = {2110.09080},
 primaryClass = {astro-ph.EP},
       adsurl = {https://ui.adsabs.harvard.edu/abs/2022A&A...658A..42P}
}

@ARTICLE{Irwin:2008,
       author = {{Irwin}, P.~G.~J. and {Teanby}, N.~A. and {de Kok}, R. and {Fletcher}, L.~N. and {Howett}, C.~J.~A. and {Tsang}, C.~C.~C. and {Wilson}, C.~F. and {Calcutt}, S.~B. and {Nixon}, C.~A. and {Parrish}, P.~D.},
        title = "{The NEMESIS planetary atmosphere radiative transfer and retrieval tool}",
      journal = {Journal of Quantitative Spectroscopy and Radiative Transfer},
         year = 2008,
        month = apr,
       volume = {109},
        pages = {1136-1150},
          doi = {10.1016/j.jqsrt.2007.11.006},
       adsurl = {https://ui.adsabs.harvard.edu/abs/2008JQSRT.109.1136I}
}

@ARTICLE{Lewis:2020,
       author = {{Lewis}, N.~K. and {Wakeford}, H.~R. and {MacDonald}, R.~J. and {Goyal}, J.~M. and {Sing}, D.~K. and {Barstow}, J. and {Powell}, D. and {Kataria}, T. and {Mishra}, I. and {Marley}, M.~S. and {Batalha}, N.~E. and {Moses}, J.~I. and {Gao}, P. and {Wilson}, T.~J. and {Chubb}, K.~L. and {Mikal-Evans}, T. and {Nikolov}, N. and {Pirzkal}, N. and {Spake}, J.~J. and {Stevenson}, K.~B. and {Valenti}, J. and {Zhang}, X.},
        title = "{Into the UV: The Atmosphere of the Hot Jupiter HAT-P-41b Revealed}",
      journal = {Astrophysical Journal Letters},
         year = 2020,
        month = oct,
       volume = {902},
       number = {1},
          eid = {L19},
        pages = {L19},
          doi = {10.3847/2041-8213/abb77f},
archivePrefix = {arXiv},
       eprint = {2010.08551},
 primaryClass = {astro-ph.EP},
       adsurl = {https://ui.adsabs.harvard.edu/abs/2020ApJ...902L..19L}
}

@ARTICLE{Rathcke:2021,
       author = {{Rathcke}, Alexander D. and {MacDonald}, Ryan J. and {Barstow}, Joanna K. and {Goyal}, Jayesh M. and {Lopez-Morales}, Mercedes and {Mendon{\c{c}}a}, Jo{\~a}o M. and {Sanz-Forcada}, Jorge and {Henry}, Gregory W. and {Sing}, David K. and {Alam}, Munazza K. and {Lewis}, Nikole K. and {Chubb}, Katy L. and {Taylor}, Jake and {Nikolov}, Nikolay and {Buchhave}, Lars A.},
        title = "{HST PanCET Program: A Complete Near-UV to Infrared Transmission Spectrum for the Hot Jupiter WASP-79b}",
      journal = {Astronomical Journal},
         year = 2021,
        month = oct,
       volume = {162},
       number = {4},
          eid = {138},
        pages = {138},
          doi = {10.3847/1538-3881/ac0e99},
archivePrefix = {arXiv},
       eprint = {2104.10688},
 primaryClass = {astro-ph.EP},
       adsurl = {https://ui.adsabs.harvard.edu/abs/2021AJ....162..138R}
}

@ARTICLE{Barstow:2022,
       author = {{Barstow}, Joanna K. and {Changeat}, Quentin and {Chubb}, Katy L. and {Cubillos}, Patricio E. and {Edwards}, Billy and {MacDonald}, Ryan J. and {Min}, Michiel and {Waldmann}, Ingo P.},
        title = "{A retrieval challenge exercise for the Ariel mission}",
      journal = {Experimental Astronomy},
         year = 2022,
        month = apr,
       volume = {53},
       number = {2},
        pages = {447-471},
          doi = {10.1007/s10686-021-09821-w},
archivePrefix = {arXiv},
       eprint = {2203.00482},
 primaryClass = {astro-ph.EP},
       adsurl = {https://ui.adsabs.harvard.edu/abs/2022ExA....53..447B}
}

@ARTICLE{Tinetti:2020,
       author = {{Tinetti}, Giovanna and {Eccleston}, Paul and {Haswell}, Carole and {Lagage}, Pierre-Olivier and {Leconte}, J{\'e}r{\'e}my and {L{\"u}ftinger}, Theresa and {Micela}, Giusi and {Min}, Michel and {Pilbratt}, G{\"o}ran and {Puig}, Ludovic and et al.},
        title = "{Ariel: Enabling planetary science across light-years}",
      journal = {ESA Ariel Mission Definition Study Report},
         year = 2020,
        month = nov,
          eid = {arXiv:2104.04824},
        pages = {arXiv:2104.04824},
          doi = {10.48550/arXiv.2104.04824},
archivePrefix = {arXiv},
       eprint = {2104.04824},
 primaryClass = {astro-ph.IM},
       adsurl = {https://ui.adsabs.harvard.edu/abs/2021arXiv210404824T}
}
\end{document}